# *p*-adic Strings Then and Now[1]

Peter G.O, Freund
Department of Physics and Enrico Fermi Institute
University of Chicago, Chicago, IL 60637

**Abstract**:

After a brief review of the idea and main results of the original *p*-adic string work, I describe the recent interest in *p*-adic strings in the context of AdS/CFT duality

---

[1] I dedicate this invited paper at the Sixth International Conference on *p*-adic Mathematical Physics and its Applications, Mexico-City, October 2017, to my friend, the distinguished physicist, Thomas Curtright on the occasion of his ecliptic 69th birthday.

It is with both interest and surprise that I revisit the *p*-adic strings, which were at the center of my work some thirty years ago. The interest comes from their relevance to our getting a better understanding of modern ideas like AdS/CFT duality, and non-local field theories. The surprise comes from the fact that in that distant past we managed to ask some good questions and, given the simplicity of *p*-adic strings, provide some robust answers.

Given the topic of this conference, I will assume that everybody here is familiar with the mathematics of *p*-adic numbers. We constructed [1, 2] the *p*-adic tree-level string amplitudes starting from the familiar Koba-Nielsen expressions for the N-point Archimedean, i.e. ordinary, tree-level string amplitudes. These are given in terms of real integration over certain combinations of kinematics-dependent multiplicative characters of the field **R** of real numbers. In these expressions we simply switched from real integration to *p*-adic integration, and from characters of **R** to characters of **Q**$_p$ the field of *p*-adic numbers. The outcome of this *p*-adic generalization was a set of meromorphic, *p*-adic-Möbius-invariant, crossing symmetric, factorizing *N*-point amplitudes free of pairs of incompatible poles, just like the Archimedean string amplitudes from which we started. It should also be pointed out that in this theory quantum amplitudes are **C**-valued as usual, and target space-time is an ordinary manifold with charts into **R**$^n$

The poles of these meromorphic *p*-adic string amplitudes are located at

$$s_N = -2 + i4\pi N/\ln p, \quad N \in \mathbb{Z}.$$

$s_0$ is the usual tachyon in the standard units used in Archimedean string theory. The other $s_N$ for $N \neq 0$, are periodic repeats of the tachyon, the amplitudes being periodic functions of the kinematic variables (which remain complex-valued) with imaginary period $i4\pi N/\ln p$. As complex poles on the physical sheet of string amplitudes, they correspond to a causality violation at a distance of order $\sqrt{\ln p}$ times the Planck length. Were one to choose a very large prime $p$, this could be quite embarrassing. But then, the very need to choose a prime, adds a parameter to the theory, without any phenomenological reason to do so. The mathematically most natural way to avoid this parameter proliferation is the adelic way [3]. It considers the amplitudes for the *p*-adic theories for *all* primes $p$ together with the Archimedean theory and establishes the connection through product formulas.

All the *N*-point amplitudes have been calculated explicitly, (for references see the review [4]) and they can be obtained from an explicitly obtained lagrangian of a 26-dimensional real scalar field theory [2]. The scalar field is, of course, the tachyon field. In the usual Archimedean case, obtaining such an effective tachyon lagrangian, by integrating out all other degrees of freedom, is a formidable task. In the *p*-adic case there are no other degrees of freedom beyond the tachyon, so the effective lagrangian is easily arrived at. Not surprisingly, this lagrangian is not local: it contains an infinite number of derivatives in the form of an exponential of the 26-D D'Alembertian acting on the scalar field. This was to be expected from the presence of the above mentioned complex poles in the *p*-adic string amplitudes. This scalar field theory has a dynamical non-topological soliton solution [2].

So far, all we have done is combine string theory with a remarkable piece of number theory. I call this remarkable because the existence of *p*-adic numbers reveals the fact that prime numbers are not only the multiplicative building blocks of all integers, but they also have a second purpose: they label all ways other than the familiar Archimedean way, in which a concept of continuity can be arrived at starting from the rational numbers. But usual string theory can be presented as a field theory on the 2-real-dimensional string world sheet, itself a Riemann surface embedded in the 26-real-dimensional target space. What is the world sheet of a $p$-adic string?

Naively one may guess that it is a 2-$p$-adic–dimensional surface embedded in the 26-real-dimensional target space. But there is no way one can embed the $p$-adic numbers in the reals or vice-versa. But then, how can we still speak of a string world sheet? A beautiful answer to this question was given by Zabrodin [5], who found that the world sheet of the $p$-adic string is a Bruhat-Tits tree. This turns out [6,7,8] to be closely related to the modern idea of Maldacena's AdS/CFT duality [9], and as such is of current interest. To get an idea if how this works, remember first that this duality connects a stringy bulk theory including gravity on an AdS background, with a conformal field theory in one fewer dimension. With our familiar SO(3,1) Lorentz group, the AdS group is SO(3,2) and the corresponding 4-dimensional AdS$_4$ space is SO(3,2)/SO(3,1).

When dealing with the string world sheet, we are in two dimensions, and in the Euclidean version the Lorentz group is SO(2), so that in this Euclidean version the AdS group is SO(2,1) and the 2-dimensional AdS$_{E2}$ space becomes SO(2,1)/SO(2). Now SO(2,1) ~ SL(2, $\mathbb{R}$) and SO(2) ~ U(1), so that AdS$_{E2}$ can be written as SL(2, $\mathbb{R}$)/U(1), or SL(2,$\mathbb{Q}_\infty$)/U(1) in a more suggestive notation. It has the mathematically familiar form G/K, where G is a non-compact Lie group and K is its maximal compact subgroup. Switching to the $p$-adic case we readily see [4] that the corresponding AdS space is PGL(2,$\mathbb{Q}_p$)/PGL(2,$\mathbb{Z}_p$). Here we have taken into account the fact that the maximal compact subgroup of PGL(2,$\mathbb{Q}_p$) is PGL(2,$\mathbb{Z}_p$) where $\mathbb{Z}_p$ is the set of *p*-adic integers.

In the Archimedean case SL(2, $\mathbb{R}$) had three continuous parameters and U(1) had one parameter, so that the AdS space ended up 2-dimensional. By contrast, in the *p*-adic case both PGL(2,$\mathbb{Q}_p$) and PGL(2,$\mathbb{Z}_p$) have the same number of continuous parameters, so that the *p*-adic AdS space ends up a discrete space: the Bruhat-Tits tree in which p+1 edges meet at every vertex and in which there are no closed paths. This tree sits comfortably in 26-D real target space, and we can hold on to the picture of a string moving in this target space.

The Bruhat-Tits tree functions as a natural discrete analog of the Archimedean AdS$_{E2}$ bulk space, while its boundary $\mathbb{Q}_p$ is the natural continuous non-Archimedean analog of $\mathbb{Q}_\infty$, the continuous real line. Thus, going *p*-adic, we naturally encounter a holographic set-up [6, 7, 8]. This fact has been successfully exploited [6, 7, 8] in calculations of bulk-to-bulk and bulk-to-boundary propagators. In the latter, taking the limit in which the bulk point approaches the boundary, one can then find the p-adic string two-point correlator. Similarly one can also study N-point functions and develop an AdS/CFT approach to the early work on *p*-adic strings. One can then try to connect with the geometric picture of Archimedean AdS space. Such a connection has been proposed, but it is not clear how to fully establish its adelic relation with the Archimedean case. But then, in the Archimedean *superstring* theory [10] which is the proper terrain for AdS/CFT duality, the product of an AdS space and a sphere of suitable dimensions has emerged as the natural background. It corresponds to a certain solution [11] of a supergravity theory which in its Bose sector, along with the Einstein-Hilbert action, contained also an antisymmetric tensor field. It is maybe desirable to understand what the analog of such an antisymmetric tensor field on a Bruhat-Tits tree should be, and whether analogs of the Archimedean supergravity solution mentioned above exist. If they do, they may restore an adelic connection.

One may also wonder what the *p*-adic counterpart of the Archimedean AdS radius is. Not surprisingly, it turns out [7] to be $a/\ln p$, with $a$ the common length of all edges of the Bruhat-Tits tree.

Finally, this 2-dimensional discrete bulk $AdS_2$ with a 1-dimensional boundary and a corresponding $CFT_1$ can be generalized [6] to an n-dimensional bulk $AdS_n$ and an n-1-dimensional $CFT_{n-1}$ by considering algebraic field extensions.

It should be mentioned[2] here that *p*-adic string theory has also been useful in work on certain remarkable string theory conjectures [13], which are very difficult to prove in the Archimedean case, but which can be quite readily checked in the *p*-adic case [14].

The fact that a discrete bulk like the Bruhat-Tits tree has a continuous boundary, the *p*-adic line, has found other interesting applications as well. In particular, it emerged as a crucial ingredient in a picture of eternal inflation [15].

It appears to me that in physics *p*-adic numbers are here to stay, vindicating yet another insight of André Weil (see the appendix).

---

[2] A different type of *p*-adic strings [12] in which target space-time *and* quantum amplitudes are also *p*-adically valued, falls outside the topics reviewed here.

**Appendix**

In nineteenth century Berlin, artistic salons were very fashionable. A particular salon of this type hosted by Friedrich August von Stägermanns required that the discussions be conducted in verse, as a "Liederspiel." This was not as farfetched as it sounds, for a number of well-known poets and writers participated, among them Wilhelm Müller, Achim von Arnim and E.T.A Hoffmann. There were also painters and sculptors, among them the fashionable portraitist Wilhelm Hensel.

Wilhelm Müller so much enjoyed this game that he wrote up the five poems he had made up for it, then wrote another twenty afterwards and published them all as a booklet under the title "Die schöne Müllerin." He gave a copy of this booklet to his friend the great composer Felix Mendelssohn Bartholdy. Soon thereafter Mendelssohn's sister Fanny married the painter Wilhelm Hensel (his other sister Rebecka married the great mathematician Peter Gustav Lejeune Dirichlet), and Felix travelled to Vienna, where he visited the great Austrian composer Franz Schubert to whom he gave as a gift Müller's "Die schöne Müllerin." Schubert set these poems in one of his most famous song cycles. A few years later he also set Müller's "Die Winterreise," in what along with Robert Schumann's "Dichterliebe," became the two greatest song cycles ever composed.

In the early 1980's I, a baritone, sang "Die Winterreise" in a concert at the Music Department of the University of Chicago. With P.D.B. Collins at the piano, if I remember correctly, I also sang excerpts from it at a Physics workshop in Seattle, and this earned both of us a marvelous free dinner prepared by John Iliopoulos. With Hermann Nicolai at the piano we performed it for our friend Bruno Zumino.

Back to the nineteenth century, Wilhelm Hensel and Fanny Mendelssohn Bartholdy had a grandson Kurt Hensel, who got his PhD in mathematics under the sponsorship of Leopold Kronecker, who transmitted to his student the ideas and results of his teacher Ernst Kummer. Hensel realized that Kummer's results could be derived much more simply once one introduced the new concept of the *p*-adic numbers.

No wonder, a few years after my "Winterreise" performances, with my student Mark Olson we introduced the *p*-adic strings. I gave a talk about them at the Institute in Princeton, and André Weil told me "I always knew that sooner or later *p*-adic numbers will appear in Physics."